\documentclass[10pt,preprint,a4paper]{aastex}

\usepackage{graphics,epsf}
\usepackage{amsmath}                
\usepackage{amsfonts}               
\usepackage{amssymb}                
\usepackage{epsfig}                 

\def \s{~\rm{s}}
\def \km{~\rm{km}}

\def \AU{~\rm{AU}}

\def \yr{~\rm{yr}}

\title{CLOSE STELLAR BINARY SYSTEMS BY GRAZING ENVELOPE EVOLUTION}

\author{Noam Soker\altaffilmark{1}}

\altaffiltext{3}{Department of Physics, Technion -- Israel
Institute of Technology, Haifa 32000 Israel;
soker@physics.technion.ac.il.}

\begin{document}

\begin{abstract}
I suggest a spiral-in process by which a stellar companion graze
the envelope of a giant star while both the orbital separation and
the giant radius shrink simultaneously, and a close binary system
is formed. The binary system might be viewed as evolving in a
constant state of `just entering a common envelope (CE) phase'. In
cases where this process takes place it can be an alternative to
the CE evolution where the secondary star is immerses in the
giant's envelope. The grazing envelope evolution (GEE) is made
possible only if the companion manages to accrete mass at a high
rate and launch jets that remove the outskirts of the giant
envelope, hence preventing the formation of a CE . The high
accretion rate is made possible by the accretion disk that
launches jets that efficiently carry the excess angular momentum
and energy from the accreted mass. The orbital decay itself is
caused by the gravitational interaction of the secondary star with
the envelope inward to its orbit, i.e., dynamical friction
(gravitational tide). Mass loss through the second Lagrangian
point can carry additional angular momentum and envelope mass. The
GEE lasts for tens to hundreds of years. The high accretion rate
with peaks lasting months to years might lead to a bright object
termed intermediate luminosity optical transient (ILOT; Red Novae;
Red Transients). A bipolar nebula and/or equatorial ring are
formed around the binary remnant.
\end{abstract}

\section{INTRODUCTION}
\label{sec:intro}

It is commonly accepted that close binary systems where at least
one of the stars is a stellar remnant, mainly a white dwarf (WD)
or a neutron star (NS), have evolved through a common envelope
(CE) phase (e.g., \citealt{Paczynski1976, vandenHeuvel1976,
IbenLivio1993, TaamSandquist2000, Podsiadlowski2001, Webbink2008,
TaamRicker2010, RickerTaam2012, Ivanovaetal2013}). There are some
major open questions with regards to the CE evolution. One of
these is the duration of the final CE phase, months (e.g.,
\citealt{SandquistTaam1998, DeMarco2003, DeMarco2009, Passy2011,
RickerTaam2012}), or maybe only days to several weeks (e.g.,
\citealt{RasioLivio1996}; \citealt{LivioSoker1988}). Another open
question involves the process that determines the final
core-secondary orbital separation.

The most common practice to calculate the final orbital separation
of the CE phase is to equate the gravitational energy released by
the spiraling-in binary system, $E_G$, to the envelope binding
energy, $E_{\rm bind}$, with an efficiency of $\alpha_{\rm CE}$:
$\alpha_{\rm CE} E_G = E_{\rm bind}$ (e.g., \citealt{Webbink1984,
TaurisDewi2004, Ivanovaetal2013}; see \citealt{NelemansTout2005}
for an alternative). Many researchers add the envelope internal
energy in the energy-balance equation of this $\alpha_{\rm
CE}$-prescription (e.g. \citealt{Han1994, Zorotovic2010, XuLi2010,
Davis2011, Rebassa-Mansergas2012, IvanovaChaichenets2011}). {{{{{
Numerical studies of the CE process have made a progress over the
years (e.g., \citealt{SandquistTaam1998, Lombardi2006,
DeMarco2011, Passy2011, Passyetal2012a, RickerTaam2012}), e.g.,
the $\alpha_{\rm CE} =0.25$ found in the study of
\cite{RickerTaam2012} agrees with the value derived from
observations of close binary systems as presented by
\cite{NordhausSpiegel2013}. Still, better numerical resolutions
and more computer resources for longer runs are required to show
that the CE process can indeed leads to the observed systems.
}}}}}

In light of some difficulties in the $\alpha_{\rm
CE}-$prescription (e.g., \citealt{Soker2013}) it has been
suggested that in many cases jets launched by the more compact
companion can facilitate envelope ejection, and that the final
spiraling-in process is by migration, i.e., interaction of the
binary system with a circumbinary thick disk or a flattened
envelope \citep{Soker2004, KashiSoker2011, Soker2013, Soker2014}.
{{{{  CE ejection by jets launched from a NS companion were
studied by \cite{ArmitageLivio2000} and \cite{Chevalier2012}, but
not as a general CE ejection process. The present approach is that
the launching of jets by the companion is a generic CE ejection
process (although not in all CE cases). }}}}

Here I raise the following question: Can it be that jets launched
by the secondary star will prevent the formation of a CE phase
altogether as the secondary star spirals-in toward the core of the
giant star? The necessary ingredients for such a grazing envelope
evolution (GEE) are discussed in section \ref{sec:GE}. I suggest
there a new type of spiraling-in process where the more compact
companion grazes the envelope of a giant star, and ejects the
outer envelope layers as it circles the giant and spirals-in. At
the heart of this process are jets, or disk winds, that are
launched by the secondary star. Energy considerations are
discussed in section \ref{sec:Energy}, {{{{ two specific examples
are discussed in section \ref{sec:examples}, }}}} and a short
summary is in section \ref{sec:summary}.

\section{INGREDIENTS OF THE GRAZING ENVELOPE EVOLUTION (GEE)}
\label{sec:GE}

There are two basic mechanisms that lead to the formation a CE
phase of a giant and a more compact secondary star. (1) The giant
transfers mass to the secondary at a rate that the secondary
cannot accommodate. The secondary inflates a large envelope that
merges with the giant envelope. (2) The giant expands and/or the
companion spirals-in and the giant engulfs the secondary star.
This mechanism is efficient when the secondary is far from
bringing the giant envelope to synchronization with the orbital
motion.

To have a GEE both these processes must be of small to moderate
size, as discussed below. It is important to keep in mind that the
GEE is different from a regular mass transfer in binary systems in
that in the GEE the system evolves in a constant state of `just
entering a CE phase'. Namely, parts of the giant envelope overflow
the Roche lobe in a large volume around the first Lagrangian point
$L_1$, but jets (or disk wind) launched by the secondary star
prevent the formation of a CE. If it hadn't been for these jets,
the system would have entered a CE phase. {{{{ Another point of
the GEE is that if it was not for the jets that are lunched by the
secondary star, the system would have enter a CE phase. Namely, if
a system is not to enter a CE phase by the CE theory, it would
also not enter a GEE. Such an example is the Red Rectangle that is
discussed in section \ref{sec:examples}. }}}}

\subsection{High accretion rate}
\label{subsec:accretion}

To prevent a CE phase during most, or all, of the evolution, the
companion must be able to accrete at a very high rate, $\ga
10^{-4} M_\odot \yr^{-1}$, without inflating a large envelope.
This implies that energy must be removed efficiently from the
accreted gas. For an accreting NS neutrino cooling allows a mass
accretion rate of $\ga 10^{-3} M_\odot \yr^{-1}$
\citep{Chevalier2012}. For a WD it seems that such a high
accretion rate is not possible, as nuclear burning will commence
and inflate an envelope.

To prevent large expansion of main sequence (MS) stars the
internal energy per unit mass of the accreted mass must be lower
than half the magnitude of the specific gravitational energy on
the stellar surface $e_{\rm acc} < 0.5 G M_1/R_1 \equiv e_G$. The
inner boundary of an accretion disk that touches the accreting
star meets this condition as the specific kinetic energy is
$e_{\rm kin} = 0.5~ e_G$. If there is a boundary layer where the
disk gas sharply spins-down to the stellar rotation, then energy
removal from the boundary layer can lead to $e_{\rm acc} <0.5~
e_G$, and more important, can lower the entropy and further
prevent much envelope expansion (e.g.,
\citealt{HjellmingTaam1991}).

 The way to remove the excess energy is by jets (or a
collimated wind) from an accretion disk, as has been suggested for
the 1837-1856 Great Eruption of $\eta$ Carinae
\citep{KashiSoker2010}. {{{{ In the process, most of the energy in
the disk is transferred to magnetic fields that by violent
reconnection eject mass. Namely, the released gravitational
accretion energy is channelled to magnetic fields and outflows
much more than to thermal energy and radiation (Shiber, S.,
Schreier, R., \& Soker, N., 2015, in preparation). }}}} Such an
accretion disk not only launches jets, but the asymmetrical
structure allows the system to accrete at the Eddington luminosity
limit or even somewhat higher. In that model the average accretion
rate by the secondary in $\eta$ Car over the 20 years of the Great
Eruption was $\sim 0.2 - 0.3 M_\odot \yr^{-1}$; the accretion rate
during periastron passages was much higher. A bipolar nebula (the
Homunculus) was formed from this activity.

A MS companion accreting from a giant envelope as it spirals-in
will not form an accretion disk \citep{Soker2014,
MacLeodRamirezRuiz2015}. To  ensure the presence of an accretion
disk the accretion must be via RLOF while the secondary is outside
the envelope before a CE phase, or at the termination of a CE when
accretion is from a circumbinary disk \citep{Soker2014}. In the
GEE accretion is via a RLOF. Even if the secondary tries to `dive'
into the envelope, the jets will eject the envelope above and
below the secondary star, practically preventing a static envelope
to be formed outside the secondary orbit. Another fraction of the
envelope is ejected through $L_2$ \citep{Livioetal1979}, and a
fraction of the envelope is accreted onto the secondary star.

\subsection{Preventing engulfment}
\label{subsec:engulf}

New results show that the giant envelope will not expand much
because of the rapid mass loss \citep{WoodsIvanova2011,
Passyetal2012b}, and this is not a route to a CE formation. A CE
phase can be initiated via a rapid spiraling-in evolution.
 To prevent a rapid spiraling-in from occurring, in addition to launching jets, the companion must
bring the envelope close to synchronization with the orbital
motion, i.e., almost co-rotation. Otherwise tidal interaction will
be strong and bring the secondary deep into the envelope in a
relatively short time. {{{{ Some departure from synchronization is
required to allow a spiraling in process (but not too rapid) due
to gravitational drag (tidal forces). This will be the case
because mass with specific angular momentum larger than the
average in the envelope is lost by mass removal from the envelope.
}}}}

 Tidal interaction becomes strong when the primary radius
becomes $R_1 \ga 0.25 a_0$, where $a_0$ is the orbital separation
\citep{Soker1996}. By the time the binary system has spiraled-in
to $a \ga R_1 $ the primary envelope must achieve  (almost)
co-rotation. The moment of inertia of the giant envelope is taken
to be $I_1 \simeq \eta M_{\rm env} R^2_1$, where $M_{\rm env}$ is
the convective envelope mass and $R_1$ is the giant radius. For
AGB stars and red supergiants $\eta \simeq 0.2-0.24$. We will
consider cases where initially $M_2 \ll M_1$. The above condition
to reach synchronization at $a \simeq R_1 \simeq 0.25~ a_0$ gives
a constraint of
\begin{equation}
M_2 \ga 0.15 \left( \frac{\eta}{0.22} \right)M_{\rm env} .
\label{eq:M2min}
\end{equation}

The system might temporarily evolve through a Darwin unstable
phase during which spiraling-in is faster. Mass accretion rate
during this phase might also be higher, as well as mass loss
through $L_2$. After accretion and removal of some envelope mass,
the system regains Darwin stability when $I_B > 3 I_1$. Here $I_B$
and $I_1$ are the moments of inertia of the binary system and the
envelope of the primary star, respectively. The system will become
Darwin stable approximately when the secondary to envelope mass
becomes $M_2/M_{\rm env} \ga 0.7 (a/R_1)^{-2}$ (for $\eta \simeq
0.22$). Evolution then slows down.

\subsection{Evolution time}
\label{subsec:time}

Evolution time is very difficult to calculate as it is determined
by mass transfer rate and tidal interaction. These are only known
crudely for these systems. Non the less, we can crudely take the
GEE timescale to be determined by the tidal and Kelvin-Helmholtz
(thermal) time-scales, $\tau_{\rm T-ev}$ and $\tau_{\rm KH-env}$,
respectively. When the envelope mass is low we crudely have
\citep{Soker2008}
\begin{equation}
\tau_{\rm KH-env} \simeq 50 \left( \frac{M_c}{0.6 M_\odot} \right)
\left( \frac{M_{\rm env}}{0.5 M_\odot} \right) \left(
\frac{L}{5000 L_\odot} \right)^{-1} \left( \frac{R_1}{200 R_\odot}
\right)^{-1} \yr,
  \label{eq:tkh2}
\end{equation}
where $M_{c}$ is the core mass and $L_1$ is the primary (giant)
stellar luminosity.

For the tidal spiral-in timescale I take from equation (6) of
\cite{Soker1996} that is based on \cite{Zahn1977} and
\cite{VerbuntPhinney1995}, assuming that the secondary is somewhat
outside the primary $a > R_1$,
\begin{eqnarray}
\tau_{\rm T-ev} \simeq 20
  \left(   \frac {a}{1.2R_1} \right)^{8}
   \left( \frac {L} {5000 L_\odot} \right)^ {-1/3}
   \left( \frac {R}{200 R_\odot} \right)^ {2/3}
  \left( \frac{M_{\rm {env}}}{0.5M_1} \right)^{-1}\nonumber \\
   \times  \left( \frac{M_{\rm {env}}}{0.5M_\odot} \right)^ {1/3}
   \left( \frac{M_2}{0.2M_1} \right)^ {-1}
   \left( \frac{\Omega_{\rm orb} -\omega_{1}}{0.1 \Omega_{\rm
   orb}}\right)^{-1} \yr,
     \label{eq:tidal}
\end{eqnarray}
where $\omega_{1}$ is the rotational angular velocity and
$\Omega_{\rm orb}$ is the orbital angular velocity.

{{{{{ To continue the spiraling-in process a strong interaction
must take place between the envelope and the secondary star. The
two processes, tidal interaction and the adjustment of the giant
envelope to mass loss, must have sufficient time to operate. The
slower process determines the spiraling-in time. }}}}}
 Overall, I crudely estimate the timescale of
the GEE phase that starts at $\sim 1 \AU$ to be
\begin{equation}
\tau_{\rm GEE} \sim 30-300 \yr \simeq  \max (\tau_{\rm KH-env},
\tau_{\rm T-ev}) .
  \label{eq:tgee}
\end{equation}
The corresponding average mass accretion rate by the secondary
star is $\dot M_2 \sim 10^{-4} - 10^{-2} M_\odot \yr^{-1}$. {{{{ I
note that the accretion rate onto the companion estimated by
\cite{RickerTaam2012} in their CE numerical simulation is $\sim
0.01 M_\odot \yr^{-1}$. So the accretion rate required by the GEE
is not as extreme as what the first impression might be. }}}}

\section{ENERGY CONSIDERATION}
\label{sec:Energy}

Let us compare the accretion energy with that of the $\alpha_{\rm
CE}$-prescription, $E_{\alpha_{\rm CE}}$ (see also
\citealt{Soker2014}). The gravitational energy released by an
accreting MS star is
\begin{equation}
E_{\rm jets} \simeq \frac {G M_2 M_{\rm acc}}{2R_2} ,
  \label{eq:ejet1}
\end{equation}
where $M_2$ and $R_2$ are the secondary stellar mass and radius,
respectively, and $M_{\rm acc}$ is the accreted mass.
 The gravitational energy released by the binary system is
\begin{equation}
E_{\alpha_{\rm CE}} = \frac {G M_{\rm core}{M_2}}{2a_{\rm final}}
\alpha_{\rm CE},
  \label{eq:alphace1}
\end{equation}
where $M_{\rm core}$ is the mass of the core of the giant star
(now the remnant), and $a_{\rm final}$ is the final core-companion
orbital separation. To liberate energy as in the $\alpha_{\rm
CE}$-prescription, the accreted mass onto the secondary star
should be
\begin{equation}
M_{\rm acc-\alpha} \simeq 0.06 \left( \frac{M_{\rm core}}{0.6
M_\sun} \right) \left( \frac{R_2}{1 R_\sun} \right) \left(
\frac{a_{\rm final}}{5 R_\sun} \right)^{-1} \left(
\frac{\alpha_{\rm CE}}{0.5} \right) M_\sun.
  \label{eq:macc1}
\end{equation}

Namely, jets launched by a secondary star that accretes $\sim 10
\%$ of its mass can play a significant, and even the major, role
in removing the giant envelope. Moreover, the jets can ejects
large parts of the envelope at very high velocities, i.e., much
above the escape speed from AGB stars. By accreting an extra mass
of $\Delta M_{\rm acc} = M_{\rm acc} - M_{\rm acc-\alpha}$ onto a
MS star, and assuming that this extra energy is channelled to
kinetic energy of the envelope, the typical outflow velocity of
the jet-ejected envelope of mass $M_{\rm en-ej}$ is
\begin{equation}
v_{\rm en-ej} \simeq 100
 \left( \frac{\Delta M_{\rm acc}}{0.05 ~M_{\rm en-ej}} \right)^{1/2}
 \km \s^{-1}.
  \label{eq:Vjej}
\end{equation}

\section{SPECIFIC EXAMPLES}
\label{sec:examples}

{{{{{ In general the constraints on the parameters for the GEE to
take place are estimated as follows. To bring the envelope to
almost synchronization with the orbital motion we require $M_2 \ga
0.15 M_{\rm env}$ (eq. \ref{eq:M2min}). As well, the system should
get into a close contact. Hence the companion should not bring the
envelope to synchronization at a too large orbital separation.
This crudely sets a limit of $M_2 \la 0.5 M_{\rm env}$. The
initial orbital separation is like that required for the system to
enter a CE phase. And most important, the companion must be a main
sequence star (or a brown dwarf), but not a WD. These constraints
leave a large parameters space for the GEE. However, there is
another important constraint, and that is that the secondary
launch two strong jets in a continues manner. If some
instabilities stop jets injection for a too long time (about an
orbital period), the system will get into a CE phase because
without jets the secondary will not be able to remove the envelop
gas that overflows its orbit. Determining the process of high-rate
mass accretion and jets launching is for futures studies. I turn
to discuss two examples. }}}}}

{{{{ \emph{The bipolar PN NGC~2346 as a possible GEE descendant.}
A possible case where a GEE could have taken place, at least part
of the time, is the PN NGC~2346. This is a bipolar PN (e.g.,
\citealt{CorradiSchwarz1995}) with a central binary star, V651
Monocerotis, having an orbital period 15.99~day
\citep{MendezNiemela1981}. \cite{IbenLivio1993}, based on a study
by \cite{IbenTutukov1993}, suggested the following evolutionary
scenario. The primary, of mass $M1 =0.4 \pm 0.05 M_\odot$, is a
degenerate helium core, while the secondary star, of mass $M_2=1.8
\pm 0.3 M_\odot$, is a main-sequence star. The orbital semimajor
axis is $34.9 R_\odot$. \cite{IbenLivio1993} suggested that the
primary overflowed its Roche lobe while on the late red giant
branch (RGB; a late case B), and the system entered a CE phase.
The initial orbital semimajor axis was $\sim 1 \AU$. During the CE
phase the binary system spirals in to its final separation. In
\cite{Soker1998} I proposed that in NGC~2346 and similar PNe, the
binary system avoid the CE phase for a large fraction of the
interaction time. This is required for the secondary to launch
jets that shape the bipolar nebula. But the binary system still
needs to spiral-in. The GEE can comprise these two properties.
}}}}

{{{{ In short, I argue that the central binary system of NGC~2346,
which is the longest of all post-CE central binary systems of PNe,
went though a GEE, at least a substantial part of the evolution
time. }}}}

{{{{ \emph{The Red Rectangle: launching Jets but avoiding GEE.}
The Red Rectangle is a bipolar nebula around a post-AGB star in a
binary system. The binary system, HD 44179, has an orbital period
of 318~days, and the secondary star is thought to be a main
sequence star of mass $\sim 0.94 M_\odot$ \citep{Waelkensetal1996,
Wittetal2009}. The companion launches two jets (e.g.,
\citealt{Wittetal2009}) that shape the nebula. The mass transfer
rate from the primary post-AGB star to the secondary is $\sim 2-5
\times 10^{-5} M_\odot \yr^{-1}$ \citep{Wittetal2009}. This binary
system does not enter a GEE as much as it does not enter a CE
phase. The reason is that the secondary is massive enough to bring
the AGB envelope to synchronization, and the secondary is actually
more massive than the primary. Mass transfer and mass loss act to
increase the orbital separation. }}}}

{{{{ Saying that, it is possible, but not necessary, that the
system went through a GEE phase in the past. In that scenario,
when the AGB star was more massive the orbital separation
decreased and the AGB radius increased to about the same value.
Strong interaction started between the two stars. If this was the
case, to avoid a CE phase with a much smaller final orbital
separation than observed, jets launched by the secondary star were
required to eject a large envelope mass. When the envelope mass
decreased enough, and the primary mass decreased below the
secondary mass, jets removed the envelope outskirts such that
further mass loss and mass transfer increased the orbital
separation and spiraling-in ceased. The present nebula was formed
during this stable phase. Mass lost in the GEE phase is at very
large distances and maybe dispersed already in the ISM. }}}}

\section{SUMMARY}
\label{sec:summary}

In this preliminary study I explored a spiral-in process by which
a stellar companion graze the envelope of a giant star and the
system does not enter a common envelope (CE) phase. It is termed
grazing envelope evolution (GEE).
 Jets launched by the secondary star are at the heart of the GEE,
and distinguish this from a regular mass transfer evolution. The
secondary star accretes mass via a RLOF mass transfer such that an
accretion disk is formed around it. Jets carry energy from the
disk, and most likely lead to accretion of low entropy gas. The
accretion through a disk allows therefore a main sequence
secondary star to accommodate the accreted mass at a rate of $\sim
10^{-4} -0.01 M_\odot \yr^{-1}$ without inflating a large envelope
(section \ref{subsec:accretion}).

In the GEE the binary system might be viewed as evolving in a
constant state of `just entering a CE phase'. The jets (or
disk-wind) launched by the accretion disk remove the envelope from
above and below the equatorial plane in the secondary vicinity,
preventing the formation of a CE phase. Without the efficiency
envelope removal by jets the system would have entered a CE phase.
{{{{ Namely, systems that enter the GEE are the same systems that
would have enter a CE phase if it was not for the efficient
envelope removal by jets. The drag force is also similar. In the
CE phase the spiraling-in is caused mainly by gravitational drag
\citep{RickerTaam2008, RickerTaam2012}. The same holds for the
GEE, where the gravitational drag is with the envelope mass
residing inward to the secondary orbit (hence might be viewed as a
tidal force). We can summarize this schematically as follows:
 (just entering a CE phase)+(efficient jets)$\rightarrow$GEE. }}}}

Another ingredient that is necessary to slow down the formation of
a CE phase is that the secondary star manages to spin-up the
envelope of the giant to almost co-rotation, hence substantially
prolonging the tidal spiral-in time scale to tens of year and
longer (section \ref{subsec:engulf}). The over all GEE duration
time is tens to hundreds of years (section \ref{subsec:time}).

{{{{{ While main sequence stars can go through the GEE, WD cannot
accrete at a high enough rate and launch jets. The reason is that
at a high accretion rate nuclear burning on the surface of the WD
inflate an envelope. Therefore, if jets aid indeed in removing the
envelope, then WDs are more likely to merge with AGB cores than
main sequence stars are. Merger of a WD with a core of an AGB star
might eventually lead to a type Ia supernovae according to the
core-degenerate scenario \citep{KashiSoker2011}.  }}}}}

The accretion of mass onto a main sequence star over tens of
years, with some bright peaks lasting months to few years of
higher than average accretion rates, can lead to a transient
object termed intermediate luminosity optical transient (ILOT; Red
Novae; Red Transients). The binary interaction and the ejection of
envelope mass by jets (section \ref{sec:Energy}) lead to a bipolar
nebula and/or equatorial ring around the binary remnant.

What is the relation of the GEE to the CE evolution? It seems that
there is a bistable evolutionary situation at hand. If the
companion enters the envelope and the accretion flow is more like
a Bondi-Hoyle-Lyttleton accretion from a medium, where an
accretion disk does not formed around a main sequence star, then
it will spiral-in faster than the ejection of the envelope. The
binary system experiences a CE evolution. If an accretion disk
that launches jets is formed before the secondary enters the
envelope, on the other hand, and the system is more or less
synchronized such that tidal spiraling-in time scale is long, the
system evolves through the GEE. If for some reason one the the
condition ceases to exist, e.g., the primary suffers instabilities
and rapidly expands to engulf the secondary star or the secondary
suffers instabilities that prevent jets from being lunched, then
the binary system jumps from the GEE to the CE phase.  {{{{ The
question of when efficient jets, namely efficient in removing the
envelope, are formed, is an question to be determined in future
studies that involve heavy 3D numerical simulations. Only then we
will be able to construct the binary system parameters space for
the GEE. }}}}



\begin{thebibliography}

\bibitem[Armitage \& Livio(2000)]{ArmitageLivio2000}  {{{{   Armitage, P.~J., \& Livio, M.\ 2000, \apj, 532,
540 }  }}}

\bibitem[Chevalier(2012)]{Chevalier2012} Chevalier, R.~A.\ 2012, \apj, 752, L2

\bibitem[Corradi \& Schwarz(1995)]{CorradiSchwarz1995}  {{{{ Corradi, R.~L.~M., \&
Schwarz, H.~E.\ 1995, \aap, 293, 871 }}}}

\bibitem[Davis et al.(2011)]{Davis2011} Davis, P.~J., Kolb, U., \& Knigge, C.\ 2011, arXiv:1106.4741

\bibitem[De Marco et al.(2009)]{DeMarco2009} De Marco, O., Farihi, J., \& Nordhaus, J.\ 2009, Journal of Physics Conference Series, 172, 012031

\bibitem[De Marco et al.(2011)]{DeMarco2011} De Marco, O., Passy, J.-C., Moe, M., Herwig, F., Mac Low, M.-M., \& Paxton, B.\ 2011, \mnras, 411, 2277

\bibitem[De Marco et al.(2003)]{DeMarco2003} De Marco, O., Sandquist, E.~L., Mac Low, M.-M., Herwig, F., \& Taam, R.~E.\ 2003, Revista Mexicana de Astronomia y Astrofisica Conference Series, 18, 24

\bibitem[Han et al.(1994)]{Han1994} Han, Z., Podsiadlowski, P., \& Eggleton, P.~P.\ 1994, \mnras, 270, 121

\bibitem[Hjellming \& Taam(1991)]{HjellmingTaam1991} Hjellming, M.~S., \& Taam,
R.~E.\ 1991, \apj, 370, 709

\bibitem[Iben \& Livio(1993)]{IbenLivio1993} Iben, I., Jr., \& Livio, M.\ 1993, \pasp, 105, 1373

\bibitem[Iben \& Tutukov(1993)]{IbenTutukov1993}  {{{{ Iben, I., Jr., \& Tutukov,
A.~V.\ 1993, \apj, 418, 343 }}}}

\bibitem[Ivanova \& Chaichenets(2011)]{IvanovaChaichenets2011} Ivanova, N., \& Chaichenets, S.\ 2011, \apjl, 731, L36

\bibitem[Ivanova et al.(2013)]{Ivanovaetal2013} Ivanova, N., Justham, S., Chen, X., et al.\ 2013, \aapr, 21, 59

\bibitem[Kashi \& Soker(2010)]{KashiSoker2010} Kashi, A., \& Soker, N.\ 2010, \apj, 723, 602

\bibitem[Kashi \& Soker(2011)]{KashiSoker2011} Kashi, A., \& Soker, N.\ 2011, \mnras, 417, 1466

\bibitem[Livio et al.(1979)]{Livioetal1979} Livio, M., Salzman, J., \& Shaviv, G.\ 1979, \mnras, 188, 1

\bibitem[Livio \& Soker(1988)]{LivioSoker1988} Livio, M., \& Soker, N.\ 1988, \apj, 329, 764

\bibitem[Lombardi et al.(2006)]{Lombardi2006} Lombardi, J.~C., Jr., Proulx, Z.~F., Dooley, K.~L., Theriault, E.~M., Ivanova, N., \& Rasio, F.~A.\ 2006, \apj, 640, 441

\bibitem[MacLeod \& Ramirez-Ruiz(2015)]{MacLeodRamirezRuiz2015} MacLeod, M.,  \& Ramirez-Ruiz, E.\ 2015,  (arXiv:1410.3823)

\bibitem[Mendez \& Niemela(1981)]{MendezNiemela1981}  {{{{ Mendez, R.~H., \& Niemela,
V.~S.\ 1981, \apj, 250, 240 }}}}

\bibitem[Nelemans \& Tout(2005)]{NelemansTout2005} Nelemans, G., \& Tout, C.~A.\
  2005, \mnras, 356, 753

\bibitem[Nordhaus \& Spiegel(2013)]{NordhausSpiegel2013} {{{{{ Nordhaus, J., \& Spiegel,
D.~S.\ 2013, \mnras, 432, 500 }}}}}

\bibitem[Paczynski(1976)]{Paczynski1976} Paczynski, B.\ 1976, Structure and Evolution of Close Binary Systems, IAU Symposium Nr.73, 75 (Reidel, Dordrecht)

\bibitem[Papish et al.(2013)]{Papishetal2013} Papish, O., Soker, N., \& Bukay, I.\ 2013, arXiv:1309.3925

\bibitem[Passy et al.(2011)]{Passy2011} Passy, J.-C., De Marco, O., Fryer, C.~L., et al.\ 2012, \apj, 744, 52

\bibitem[Passy et al.(2012a)]{Passyetal2012a} Passy, J.-C., De Marco, O., Fryer, C.~L., et al.\ 2012a, \apj, 744, 52

\bibitem[Passy et al.(2012b)]{Passyetal2012b} Passy, J.-C., Herwig,
F., \& Paxton, B.\ 2012b, \apj, 760, 90

\bibitem[Podsiadlowski(2001)]{Podsiadlowski2001} Podsiadlowski, P.\ 2001, Evolution of Binary and Multiple Star Systems, 229, 239

\bibitem[Rasio \& Livio(1996)]{RasioLivio1996} Rasio, F.~A., \& Livio, M.\ 1996, \apj, 471, 366

\bibitem[Rebassa-Mansergas et al.(2012)]{Rebassa-Mansergas2012} Rebassa-Mansergas, A., Zorotovic, M., Schreiber, M.~R., et al.\ 2012, arXiv:1203.1208

\bibitem[Ricker \& Taam(2008)]{RickerTaam2008}  {{{{  Ricker, P.~M., \& Taam, R.~E.\
2008, \apjl, 672, L41 }}}}

\bibitem[Ricker \& Taam(2012)]{RickerTaam2012} Ricker, P.~M., \& Taam, R.~E.\ 2012, \apj, 746, 74

\bibitem[Sandquist et al.(1998)]{SandquistTaam1998} Sandquist, E.~L., Taam, R.~E., Chen, X., Bodenheimer, P., \& Burkert, A.\ 1998, \apj, 500, 909

\bibitem[Soker(1996)]{Soker1996} Soker, N.\ 1996, \apjl, 460, L53

\bibitem[Soker(1998)]{Soker1998}  {{{{ Soker, N.\ 1998, \apj, 496,
833}}}}

\bibitem[Soker(2004)]{Soker2004} Soker, N.\ 2004, \na, 9, 399

\bibitem[Soker(2008)]{Soker2008} Soker, N.\ 2008, \apjl, 674, L49

\bibitem[Soker(2013)]{Soker2013} Soker, N.\ 2013, \na, 18, 18

\bibitem[Soker(2014)]{Soker2014} Soker, N.\ 2014, arXiv:1404.5234

\bibitem[Taam \& Ricker(2010)]{TaamRicker2010} Taam, R.~E., \& Ricker, P.~M.\ 2010, \nar, 54, 65

\bibitem[Taam \& Sandquist(2000)]{TaamSandquist2000} Taam, R.~E., \& Sandquist, E.~L.\ 2000, \araa, 38, 113

\bibitem[Tauris \& Dewi(2001)]{TaurisDewi2004} Tauris, T.~M., \& Dewi, J.~D.~M.\ 2001, \aap, 369, 170

\bibitem[van den Heuvel(1976)]{vandenHeuvel1976} van den Heuvel, E.~P.~J.\ 1976, in IAU Symposium, Vol. 73,
      Structure and Evolution of Close Binary Systems, ed. P. Eggleton, S. Mitton, \& J. Whelan, 35

\bibitem[Verbunt \& Phinney(1995)]{VerbuntPhinney1995} Verbunt, F., \& Phinney,
E.~S.\ 1995, \aap, 296, 709

\bibitem[Waelkens et al.(1996)]{Waelkensetal1996}  {{{{ Waelkens, C., Van Winckel, H.,
Waters, L.~B.~F.~M., \& Bakker, E.~J.\ 1996, \aap, 314, L17 }}}}

\bibitem[Webbink(1984)]{Webbink1984} Webbink, R.~F.\ 1984, \apj, 277, 355

\bibitem[Webbink(2008)]{Webbink2008} Webbink, R.~F.\ 2008, Astrophysics and Space Science Library, 352, 233

\bibitem[Witt et al.(2009)]{Wittetal2009}  {{{{ Witt, A.~N., Vijh, U.~P.,
Hobbs, L.~M., Aufdenberg, J. P., Thorburn, J.~A.; York, D.~G.\
2009, \apj, 693, 1946 }}}}

\bibitem[Woods \& Ivanova(2011)]{WoodsIvanova2011} Woods, T.~E., \& Ivanova,
N.\ 2011, \apjl, 739, L48

\bibitem[Xu \& Li(2010)]{XuLi2010} Xu, X.-J., \& Li, X.-D.\ 2010, \apj, 716, 114

\bibitem[Zahn(1977)]{Zahn1977} Zahn, J.-P.\ 1977, \aap, 57, 383

\bibitem[Zorotovic et al.(2010)]{Zorotovic2010} Zorotovic, M., Schreiber, M.~R., G{\"a}nsicke, B.~T., \& Nebot G{\'o}mez-Mor{\'a}n, A.\ 2010, \aap, 520, A86

%


\end{thebibliography}
\end{document}